\documentclass[aps,prd,amsfonts,preprint, eqsecnum,nofootinbib,axodraw]
{revtex4} 
\usepackage{graphics,epsfig}
\begin{document}
\def\kkdl{d_L^\bullet} 
\def\kkdr{d_R^\bullet} 
\def\kkul{u_L^\bullet} 
\def\kkur{u_R^\bullet} 
\def\kkw{W^\bullet} 
\def\kkz{Z^\bullet} 
\def\kknu{\nu^\bullet} 
\def\kkl{l^\bullet} 
\def\kkb{B^\bullet} 
\def\kkdlbar{\bar{d}_L^\bullet} 
\def\kkdrbar{\bar{d}_R^\bullet} 
\def\kkulbar{\bar{u}_L^\bullet} 
\def\kkurbar{\bar{u}_R^\bullet} 
\def\la{\langle}
\def\ra{\rangle}
\def\CPTslash{{\small CPT}{\kern -2.0ex\hbox{/}}}
\def\etal {{\it et al.}}

\preprint{WM-04-102}
%

%
\title{\vspace*{0.5in}Lorentz and CPT Violation in the Higgs Sector \vskip 0.1in}

\vskip .8in
\def\bar{\overline}

\author{David L. Anderson}\email[]{dlande@wm.edu}
\author{Marc Sher}\email[]{sher@physics.wm.edu}
\author{Ismail Turan}\email[]{ituran@newton.physics.metu.edu.tr} 
\affiliation{Particle Theory Group, Department of Physics,
College of William and Mary, Williamsburg, VA 23187-8795}
\date{March 2004}

\begin{abstract}
{\narrower\narrower  Colladay and Kosteleck\'y have proposed a 
framework for studying Lorentz and CPT violation in a natural extension of the Standard 
Model.  Although numerous bounds exist on the Lorentz and CPT 
violating parameters in the gauge boson and fermion sectors, there are 
no published bounds on the parameters in the Higgs sector.   We 
determine these bounds.   The bounds on the CPT-even asymmetric 
coefficients arise from the one-loop contributions to the photon 
propagator, those from the CPT-even symmetric coefficients arise from 
the equivalent $c_{\mu\nu}$ coefficients in the fermion sector, and 
those from the CPT-odd coefficient arise from bounds on the vacuum 
expectation value of the $Z$-boson.}
\end{abstract}
\maketitle
\section{Introduction}\label{sec:intro}
The scale of the unification of gravity with the other interactions is 
expected to be near the Planck scale of $10^{19}$ GeV.   This is far 
out of reach of any future accelerators and thus is not directly 
experimentally accessible.   However, the nonlocality of string 
theory leads to the possibility that Lorentz and CPT symmetry violations 
might exist at that scale \cite{kostelecky:1991ak}, and hence high-precision studies of these 
symmetries might be able to probe Planck-scale physics.

It is difficult to write the most general Lorentz and CPT violating 
theory--even the meaning of a Lagrangian becomes questionable in such 
a theory. However, with some reasonable assumptions, one can 
study Lorentz and CPT violation. To develop a 
framework for studying Lorentz and CPT 
violation in the Standard Model, Colladay and Kosteleck\'y \cite{Colladay:1998fq} 
constructed the Standard Model Extension (SME). This is a theory based 
on the standard model but which includes additional Lorentz and CPT 
violating terms. These terms satisfy the $\rm SU(3)\times SU(2)\times 
U(1)$ gauge symmetry of the Standard Model, and they also satisfy 
invariance under observer Lorentz transformations \cite{Colladay:1998fq,Lehnert:2003ue,Colladay:1996iz}. 
This means that any Lorentz indices that the additional term contains 
must be contracted (i.e., it must be an observer Lorentz scalar), and that rotations and 
boosts of 
the observer inertial frame do not affect the physics. This ensures 
that the physics does not depend on the choice of coordinates.  In 
addition, the Lorentz violation is assumed independent of position 
and time, 
and thus energy and momentum are conserved.  The Lorentz-violating terms considered in the SME violate invariance under particle Lorentz 
transformations, i.e. under rotations and boost of a particle within 
a fixed observer inertial frame.   An example of two such terms in the 
pure electron sector is $\overline{\psi}M\psi$, where $M\equiv 
a_{\mu}\gamma^{\mu} + b_{\mu}\gamma^{\mu}\gamma_{5}$.  This term is 
clearly $\rm SU(3)\times SU(2)\times U(1)$ invariant, and the 
coefficients are position-independent, but $a_{\mu}$ and $b_{\mu}$ are 
constant vectors and do not transform under a particle Lorentz transformation. It should be noted that this is the ``minimal'' extension. Non-Minkowski spacetimes \cite{Kostelecky:2003fs} will lead to spacetime-dependent coefficients, and some models can lead to nonrenormalizable terms. Such minimal extensions are beyond the scope of this paper.

In the SME, the additional terms in the Higgs sector are given 
by \cite{Colladay:1998fq}
\begin{equation}
    {\cal L}_{\rm CPT-even}= \left[\frac{1}{2}(k_{\phi\phi}^{S}+ik_{\phi\phi}^{
    A})_{\mu\nu}(D^{\mu}\Phi)^{\dagger}D^{\nu}\Phi +{\rm H.c.}\right]-\frac{1}{2} k_{\phi 
    B}^{\mu\nu}\Phi^{\dagger}\Phi B_{\mu\nu}-\frac{1}{2}k_{\phi 
    W}^{\mu\nu}\Phi^{\dagger}W_{\mu\nu}\Phi\,,\label{LagLV}
    \end{equation}
and 
\begin{equation}
    {\cal L}_{\rm CPT-odd} = 
    ik_{\phi}^{\mu}\Phi^{\dagger}D_{\mu}\Phi + {\rm H.c.}\label{LagCPTodd}
    \end{equation}
Here, we have broken the $k_{\phi\phi}$ term up into its real 
symmetric and imaginary antisymmetric parts. Note that the $k_{\phi B}$ and 
$k_{\phi W}$ coefficients are real antisymmetric, the CPT even 
coefficients are all dimensionless, and the complex-valued CPT odd coefficient has 
units of mass.

To our knowledge, there are no published limits on the possible 
values of these coefficients.  The purpose of this article is to explore 
the current bounds on these terms.   In section II, we consider the bounds on the CPT-even antisymmetric 
coefficients, $k_{\phi\phi}^A, k_{\phi B}$ and $k_{\phi W}$.  In section III, the bounds of the CPT-even
symmetric coefficients $k_{\phi\phi}^S$ are determined, and the bounds on the CPT-odd coefficient,
$k_{\phi}$ are discussed in section IV.  Section V contains our conclusions and a summary of the bounds.

\section{Bounds on the CPT-even antisymmetric coefficients}

Whenever new particles or new interactions are proposed, there are two 
approaches to discovery.  One can look for direct detection of these 
particles or interactions (as in searches for supersymmetric particles 
or for flavor-changing neutral currents).   Alternatively, one can 
look at the loop effects of the new physics on lower energy processes, such 
as in precision electroweak measurements.  In studying the above 
coefficients, direct detection 
would necessitate producing large numbers of Higgs bosons, and the 
resulting bounds would be quite weak.  However, there are extremely 
stringent bounds on Lorentz violation at low energies, and 
thus searching for the effects of these new interactions through loop 
effects will provide the strongest bounds. The most promising of these effects 
will be on the photon propagator.

In this section, we will consider the bounds on the CPT-even antisymmetric 
coefficients, $k_{\phi\phi}^A, k_{\phi B}$ and $k_{\phi W}$.   These interactions will
lead to modified vertices and propagators, and will thus affect the one-loop photon 
propagator.  We first look at the most general CPT-even photon propagator, and then relate
the $k_{\phi\phi}^A$ coefficients to the Lorentz-violating terms in the photon propagator.  Then, 
the experimental constraints on such terms lead directly to stringent bounds on the $k_{\phi\phi}^A$ 
coefficients.  We then consider the $k_{\phi B}$ and $k_{\phi W}$ coefficients.

Considering CPT-even terms only, the photon Lagrangian can be written 
as \cite{Colladay:1998fq}
\begin{equation}
    {\cal L}_{\rm photon}=-{1\over 4}F_{\mu\nu}F^{\mu\nu} -{1\over 
    4}(k_{F})_{\kappa\lambda\mu\nu}F^{\kappa\lambda}F^{\mu\nu}\,.
    \end{equation}
    Here $k_F$ has the symmetries of the Riemann tensor plus a 
    double-traceless constraint, giving 19 independent parameters.
The equation of motion from this Lagrangian is \begin{equation} 
M^{\alpha\delta}A_{\delta}=0\,,\end{equation} where
\begin{equation}
M^{\alpha\delta}(p) \equiv g^{\alpha\delta}p^{2}-p^{\alpha}p^{\delta}-2
(k_{F})^{\alpha\beta\gamma\delta}p_{\beta}p_{\gamma}\,.\label{photprop}
\end{equation}
The propagator is clearly gauge invariant (recall that $k_{F}$ is 
antisymmetric under exchange of the first or last two indices).

To bound the coefficients, we calculate the vacuum polarization 
diagrams for the photon propagator, using the full Lagrangian, 
including Lorentz-violating terms.  The result will be of the form of 
the above propagator, and one can read off the value of $k_{F}$.  
Note that while the $g^{\mu\nu}p^{2}-p^{\mu}p^{\nu}$ structure is 
mandated by gauge invariance, the $k_{F}$ term is separately gauge 
invariant and may differ order by order in perturbation theory. For simplicity, we look at the divergent parts of the one loop diagrams only\footnote{At extremely high energies, either energy 
positivity or microcausality may be lost \cite{Kostelecky:2000mm}. However 
if we cut off 
the theory at a high, but finite, scale, this will not be an issue.}.  
Consideration of higher orders and finite 
parts will give similar, although not necessarily identical, results.

In general, due to the large number of Lorentz-violating terms, this 
yields a bound 
in a multidimensional parameter space.  However, if we do not consider 
the possibility of fine-tuning, then we can consider each of the 
possible terms independently.  One must keep in mind that some of the 
parameters may be related by a symmetry, but absent such a symmetry, 
we expect no high-precision cancellations.  We begin by considering 
the antisymmetric part of $k_{\phi\phi}$, and then $k_{\phi B}$ and 
$k_{\phi W}$.

\begin{figure}[htb]
\vskip -6.5cm
\centerline{ \epsfxsize 6.5in {\epsfbox{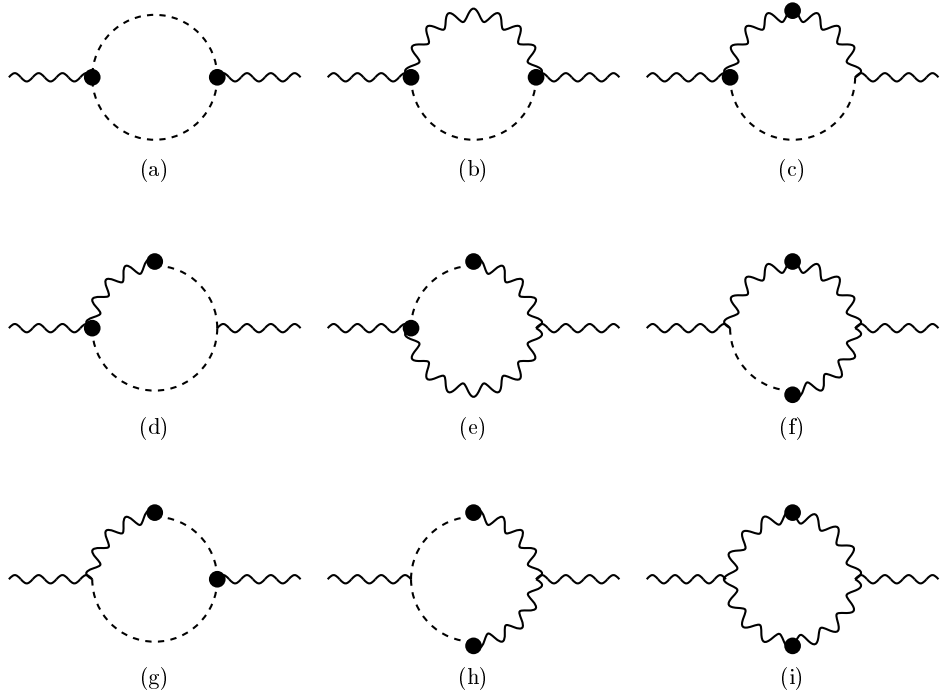}}}
        \vskip -7.8cm
\caption{One-loop contributions to the photon vacuum polarization
involving Lorentz-violating
interactions to second order. These diagrams are for
$k_{\phi\phi}^A$ case but similar diagrams
exist for the other antisymmetric coefficients. Here the wavy (dashed)
line circulating in the loop represents $W$ boson (charged Goldstone
boson).
Each blob in vertices, $W$-propagator or $W\!\!-\!\!\phi$ mixed
propagator represents a single Lorentz-violating coefficient insertion. The rest of the diagrams can be obtained by
permutations of these 9 diagrams.}\label{oneloop}\end{figure}

To calculate the additional vacuum polarization diagrams for the photon propagator 
due to a non-zero $k_{\phi\phi}^{A}$-term in Eq. (\ref{LagLV}) (assuming all other parameters are zero), 
we need to find the vertices and propagators which are dependent on $k_{\phi\phi}^{A}$. For our purpose, 
vertices involving at least one photon field are necessary. Two of them, for instance, can be quoted here: 
The $A_{\mu}W^{-}_{\nu}\phi^{+}\left[A_{\mu}(p)\phi^+\phi^-\right]$ 
coupling is given by $-e m_{W}(k_{\phi
\phi}^A)_{\mu\nu}\left[-e (k_{\phi\phi}^A)_{\mu\nu}p^{\nu}\right]$. Here 
all momenta are taken towards the vertex,
and $\phi^{\pm}$ is the usual charged Goldstone boson.
As in the conventional SM,  one can choose  acceptable 
gauge-fixing conditions to remove the redundant degrees of freedom from the theory. In the SM, the following conditions 
in the $R_{\xi}$-gauge can be chosen \cite{ChengLi}
$f_{i}=\partial_{\mu}A_{i}^{\mu}+\frac{ig\xi}{2}\left(\Phi^{\prime \dagger}\tau_i\la\Phi\ra_{0}-
\la\Phi^{\dagger}\ra_{0}\tau_i\Phi^{\prime}\right),\,\,i=1,2,3$ for $SU(2)$ case and 
$f=\partial_{\mu}B^{\mu}+\frac{ig^{\prime}\xi}{2}\left(\Phi^{\prime \dagger}\la\Phi\ra_{0}-\la
\Phi^{\dagger}\ra_{0}\Phi^{\prime}\right)$ for $U(1)$ case, where $g(g^{\prime})$ is the $SU(2)(U(1))$ coupling 
constant, $\tau_i$ are the Pauli matrices, and $\Phi^{\prime}$ and 
$\langle\Phi\rangle_{0}$ are the Higgs doublet and vacuum expectation 
value, respectively. Then the gauge-fixing term in the Lagrangian is ${\cal L}_{gf}=
-({\bf f\cdot f})^2/{2\xi}-f^2/{2\xi}$ and this removes the mixing term between $W^{\pm}$ and $\phi^{\mp}$. In the SME, 
we have additional mixing proportional to $k_{\phi\phi}^{A}$. A simple generalization of the above gauge-fixing 
conditions, by adding a $i(k_{\phi\phi}^{A})_{\mu\nu}\partial^{\mu}A_i^{\nu}$ term to $f_i$ and a similar
$i(k_{\phi\phi}^{A})_{\mu\nu}\partial^{\mu}B^{\nu}$ to the function $f$, 
would remove such Lorentz-violating mixing in 
our case as well. However, such generalization also leads to an unwanted mixing between the gauge boson $Z_{\mu}$ and 
the derivative of the Higgs field, $\partial_{\nu}\phi_1$, which is contracted 
with $(k_{\phi\phi}^{A})^{\mu\nu}$, as well as substantially 
complicating the photon propagator. 
Instead we use a mixed propagator
of the form $m_W(k_{\phi\phi}^{A})_{\mu\nu}q^{\nu}$ for $W^{\pm}_{\mu}(q)\phi^{\mp}$ fields (that is, we are treating 
the mixing term as an interaction, which leads to diagrams like 
(d),(e),(g), 
and (h) in Fig. \ref{oneloop}). Here we use the 
convention that the 4-momentum $q$ of $W_{\mu}$ is incoming to the point 
where the field turns into a charged Goldstone boson. 

Another distinct feature of this model is the presence of a term of the form $im_W(k_{\phi\phi}^{A})^{\mu\nu}W^{+}_{\mu}W^{-}_{\nu}$. 
This term needs to be considered carefully. It obviously represents a new term in the W-propagator. 
We will discuss how to deal with this term in the $R_\xi$-gauge, although we use 
't Hooft-Feynman gauge ($\xi\!\!=\!\!1$) in our vacuum polarization 
calculations. Since this mixing term can be considered  an 
interaction, one can carry out the Dyson summation. If we pick up the quadratic 
terms in the W-boson from 
the Lagrangian together with ${\cal L}_{gf}$, we have $\Delta{\cal L}^{(2)}_{W}=W_{\mu}^-K^{\mu\nu}(q)W_{\nu}^+$, 
where $iK^{\mu\nu}(q)\equiv i\left[-(q^2-m_W^2)g^{\mu\nu}+(1-1/\xi)q^{\mu}q^{\nu}+im_W^2(k_{\phi\phi}^{
A})^{\mu\nu}\right]\equiv iK^{(0)\mu\nu}(q)-m_W^2(k_{\phi\phi}^{A})^{\mu\nu}$. We know that the inverse of 
$iK^{(0)\mu\nu}(q)$, say $i\Delta_{(0)\nu\lambda}(q)$ (that is, $K^{(0)\mu\nu}\Delta_{(0)\nu\lambda}=g^\mu_{\lambda}$),
is the usual propagator for the W boson. From  $K^{\mu\nu}(q)$, one can 
write the form of the propagator as $\Delta_{\nu\lambda}(q)\equiv\Delta_{\nu\lambda}^{(0)}
(q)+B_{\nu\lambda}\left(k^A_{\phi\phi}\right)$, where all $k^A_{\phi\phi}$ 
dependence is in the second term. To determine $B_{\nu\lambda}$, we can use the fact that
 $\Delta_{\nu\lambda}$ is the inverse of $K^{\mu\nu}$. From this equation, one gets $B_{\nu\lambda}=-im_W^2\Delta^{
 (0)}_{\nu\lambda^{\prime}}(k^A_{\phi\phi})^{\lambda^{\prime}\mu}\left[\Delta^{(0)}_{\mu\lambda}+B_{\mu\lambda}\right]$.
 Iterating this equation, one obtains a series. However, we know that 
 $k^A_{\phi\phi}$ parameters are small, so it is sufficient to keep the first few terms. Up to second order, it is 
 straightforward to show that $B_{\nu\lambda}=-im_W^2 \Delta^{(0)}_{\nu\alpha}(k^A_{\phi\phi})^{\alpha\beta}\Delta^{
 (0)}_{\beta\lambda}-m_W^4\Delta^{(0)}_{\nu\alpha}(k^A_{\phi\phi})^{\alpha\alpha^{\prime}}\Delta^{
 (0)}_{\alpha^{\prime}\beta^{\prime}}(k^A_{\phi\phi})^{\beta^{\prime}\beta}\Delta^{(0)}_{\beta\lambda}$. 
 In the 't Hooft-Feynman gauge the propagator has a simple form which can be given as
\begin{eqnarray}    
i\Delta_{\nu\lambda}(\xi=1)&=&i\Delta^{(0)}_{\nu\lambda}+m_W^2\frac{(k^A_{\phi\phi})_{\nu\lambda}}{(q^2-m_W^2)^2}+im_W^4
\frac{(k^A_{\phi\phi})_{\nu\alpha}{(k^A_{\phi\phi})^{\alpha}}_{\lambda}}{(q^2-m_W^2)^3}\,,
\label{propW}
\end{eqnarray}
where, for example, the second term is represented as a blob in the $W$-propagator in Fig. \ref{oneloop}(c), Fig. \ref{oneloop}(f), and Fig. 
\ref{oneloop}(i).

We are now ready to calculate the vacuum polarization diagrams for 
the photon propagator. It is useful to classify contributions 
as the ones having first order $k^A_{\phi\phi}$-dependence and the ones with quadratic in $k^A_{\phi\phi}$. The only possible 
structure in first order is $(k^A_{\phi\phi})_{\mu\nu}$ where $\mu(\nu)$ is the Lorentz index of the incoming(outgoing) 
photon field. If we add all possible one-loop diagrams, the first order 
contributions vanish. This is expected from the gauge 
invariance requirement. 
It is not difficult to show that getting a gauge invariant 
transverse structure is only possible with 
at least two $k^A_{\phi\phi}$-terms. In Fig. \ref{oneloop}, we 
depict the one-loop diagrams which, when permutations are added, give second order 
Lorentz-violating inclusions. There are two possible structures in  second 
order, which are either 
$(k^A_{\phi\phi})_{\mu\lambda}{(k^A_{\phi\phi})^{\lambda}}_{\nu}$ or $(k^A_{\phi\phi})_{\mu\lambda}
(k^A_{\phi\phi})_{\lambda^{\prime}\nu}p^{\lambda}p^{\lambda^{\prime}}$. Here $p$ is the four momentum 
of the external photons. Again the first possibility is not gauge invariant 
and  should vanish, thus contributions from the third term in Eq. 
(\ref{propW}) should vanish.  We have verified this explicitly.  The 
latter 
is gauge 
invariant and gives a non-zero contribution (if we contract with any of two external momenta of 
photons, $p^{\mu}$ or $p^{\nu}$, it vanishes due to the antisymmetry property of $k^{A}_{\phi\phi}$). 

Calculating the one-loop diagrams, and  comparing
with Eq. (\ref{photprop}), we find that the components of $k_F$ can simply be expressed in terms of 
$k^A_{\phi\phi}$ as 
$(k_F)_{\mu\lambda\lambda^{\prime}\nu}=\frac{1}{3}(k^A_{\phi\phi})_{\mu\lambda}
(k^A_{\phi\phi})_{\lambda^{\prime}\nu}$.   
We now turn to the experimental bounds on the $k_{F}$.

The dimensionless coefficient $(k_F)_{\kappa \lambda \mu \nu}$ has the symmetries of the Riemann 
tensor and a vanishing
 double trace, resulting in nineteen independent elements.
Following Kosteleck\'y and Mewes \cite{kostelecky:2002hh}, we can 
express these elements in terms of four traceless $3 \times 3$ matrices and one coefficient:
\begin{eqnarray}
&& \left( \tilde{\kappa}_{e+} \right)^{jk} = {1 \over 2} \left( \kappa_{DE} + \kappa_{HB} \right)^{jk}, \nonumber \\
&& \left( \tilde{\kappa}_{e-} \right)^{jk} = {1 \over 2} \left( \kappa_{DE} - \kappa_{HB} \right)^{jk} - {1 \over 3}
 \delta^{ij} \left( \kappa_{DE} \right)^{ll}, \nonumber \\
&& \left( \tilde{\kappa}_{o+} \right)^{jk} = {1 \over 2} \left( \kappa_{DB} + \kappa_{HE} \right)^{jk}, \nonumber \\
&& \left( \tilde{\kappa}_{o-} \right)^{jk} = {1 \over 2} \left( \kappa_{DB} - \kappa_{HE} \right)^{jk}, \nonumber \\
&& \tilde{\kappa}_{tr} = {1 \over 3} \left( \kappa_{DE} \right)^{ll}.
\label{kappatilde}
\end{eqnarray}
where
\begin{eqnarray}
&& \left( \kappa_{DE} \right)^{jk} = -2 (k_F)^{0 j 0 k}, \nonumber \\
&& \left( \kappa_{HB} \right)^{jk} = {1 \over 2} \epsilon^{jpq} \epsilon^{krs} (k_F)^{pqrs}, \nonumber \\
&& \left( \kappa_{DB} \right)^{jk} = -\left( \kappa_{HE} \right)^{kj} = (k_F)^{0jpq}\epsilon^{kpq}.
\label{kappa}
\end{eqnarray}

There are stringent astrophysical bounds on 10 of the 19 elements, 
those given by $\tilde{\kappa}_{e+}$ and by $\tilde{\kappa}_{o-}$.
These astrophysical bounds have been discussed 
recently in detail
by Kosteleck\'y and Mewes \cite{kostelecky:2002hh}.  The 
observations of radiation propagating in free space over astrophysical distances 
results in bounds on these elements from  velocity and birefringence constraints \cite{hough,Colladay:1996iz,kostelecky:2001mb,Carroll:vb,Wolf:2004qs}. 
The bound from 
birefringence constraints is the strongest, and is given by $3\times 
10^{-32}$.  The bounds on the remaining 9 elements are much weaker 
(and in fact can be moved into the fermion sector, as will be 
discussed below).

If one of our coefficients is nonzero, say 
$(k^{A}_{\phi\phi})_{01}=-(k^{A}_{\phi\phi})_{10}\equiv x$, then the only nonzero components of 
$k_{F}$ are the $(k_F)_{1010},(k_F)_{0101},(k_F)_{1001}$ and $(k_F)_{0110}$ components.  This leads 
to a nonzero $\tilde{\kappa}_{e+}$ matrix, and thus the stringent 
bounds apply.   Extending this one can see that for any single or possible combination 
of non-zero elements of $(k_{\phi\phi}^A)_{\mu \nu}$ it is impossible 
for both
 $\tilde{\kappa}_{e+}$ and $\tilde{\kappa}_{o-}$ to be null matrices, 
 and thus the birefringence constraints apply.
 
 One cautionary note should be added.  In the above example, the $k_{F}$ tensor is not double traceless, 
 since $(k_{F})^{\mu\nu}_{\mu\nu}$ is proportional to $x^{2}$.  This 
 means that the kinetic energy for the photon has not been properly 
 normalized.  By adding and subtracting a term proportional to the double trace
 \begin{equation}
{\cal L} = - {1 \over 4} \left( 1 + \varsigma x^{2} \right) F_{\mu \nu} F^{\mu \nu}
 - {1 \over 4} (k_F)_{\kappa' \lambda' \mu' \nu'} F^{\kappa' \lambda'} F^{\mu' \nu'}
 + {1 \over 4} \left( \varsigma x^{2} \right) F_{\mu \nu} F^{\mu \nu},
\label{boundredef}
\end{equation}
where $\varsigma$  is a constant and the primed indices are summed 
only over the nonzero elements (in the above example, only over 
$(k_F)_{1010},(k_F)_{0101},(k_F)_{1001},(k_F)_{0110}$).  A redefinition of the photon field will give 
a conventional kinetic term, and the remaining terms obey the double 
traceless condition if one chooses a suitable $\varsigma$ value.  This means that, although we started with only 
a $(k_{F})_{0101}$ term (plus permutations), we also have 
$(k_F)_{0202},(k_F)_{0303},(k_F)_{1212},(k_F)_{1313}$ and $(k_F)_{2323}$ terms (plus permutations).  
Nonetheless it will still not be possible for the elements 
 of $\tilde{\kappa}_{e+}$ and $\tilde{\kappa}_{o-}$ to become zero,
hence these redefinitions do not affect the bounds. From these results, we find an upper 
bound of $3\times 10^{-16}$ for the $k^A_{\phi\phi}$ coefficients, 
barring, of course, fine-tuned cancellations.

Next, we consider the $k_{\phi B}$-term by setting all other parameters 
to 
zero in Eq. (\ref{LagLV}). This term has a interesting new 
interaction $A_{\mu}\phi_1\phi_1$, where $\phi_{1}$ is the Standard 
Model Higgs boson. There also exists a similar Lorentz-violating vertex 
with the neutral Goldstone boson, 
$\phi_2$. Therefore, in addition to the charged 
Goldstone loop, we have diagrams like Fig. 
\ref{oneloop}(a), which are second order 
in $k_{\phi B}$ with different vertex factors, where now the particles 
circulating in the loop are the 
Higgs and the would-be Goldstone bosons. The coupling is $\cos{\theta_W} 
(k_{\phi 
B})_{\mu\nu}p^{\nu}$, where $p$ is the four momentum of the photon. 
Unlike the $k^A_{\phi\phi}$ case, we obviously don't have an additional 
mixing between the $W$ and charged Goldstone bosons (thus, no diagrams 
like (d),(e),(g), and (h) in Fig. \ref{oneloop}). But this new term 
induces a remarkable mixing between the photon and the Higgs scalar, 
since when the Higgs gets a vacuum expectation value, an 
$A_{\mu}\partial_{\nu}\phi$ mixing term appears.   This term can't be 
removed by gauge-fixing, and represents a mixed propagator.
In our one-loop calculation of the photon propagator, however, the 
mixing will not contribute to the divergent part, and is thus not 
relevant\footnote{With the use of this mixing, there is an another place 
where the Lorentz-violating $k_{\phi B}$ term could contribute, 
namely in the  $A_{\mu}\bar{e}e$ and $\phi_{1}\bar{e}e$ effective vertices.  However, 
the bounds we obtain below render any such effects negligible.}. Therefore, if we look at the structures 
in the first and the second order in $k_{\phi B}$, there exist $(k_{\phi B})_{\mu\lambda}p^{\lambda}p_{\nu}$,
$(k_{\phi B})_{\nu\lambda}p^{\lambda}p_{\mu}$, and $(k_{\phi B})_{\mu\lambda}(k_{\phi B})_{\lambda^{\prime}\nu}
p^{\lambda}p^{\lambda^{\prime}}$. Note that only the scalar loop diagrams 
with two Lorentz-violating vertices 
yields the last structure (three scalar loop diagrams with charged 
Goldstone $\phi^{\pm}$, Higgs boson $\phi_1$, and would-be 
neutral Goldstone 
boson $\phi_2$). Gauge 
invariance makes us expect that the first two non-invariant structures 
should vanish and this is indeed the case. So, in this framework, the
$(k_F)_{\mu\lambda\lambda^{\prime}\nu}=
\frac{5}{12e^2}\cos^2{\theta_W}(k_{\phi B})_{\mu\lambda}(k_{\phi 
B})_{\lambda^{\prime}\nu}$ equality holds. Numerically, the 
bound on the individual $k_{\phi B}$ is stronger than that for 
$k^{A}_{\phi\phi}$ by a factor of 
$(5\cos\theta_{W}^{2}/4e^{2})^{1/2}\sim 3.2$.  This gives 
the  upper bound on $k_{\phi B}$ of $0.9\times 10^{-16}$.     

The $k_{\phi W}$ term has very similar features to the $k_{\phi\phi}^A$ case 
except for the photon-Higgs mixing. It 
additionally allows the Lorentz-violating $A_{\mu}(p)\phi_1\phi_1$ vertex, 
which is equal to $-\sin{\theta_W}k_{\mu\nu}p^{\nu}$ (leading to diagrams 
like Fig. \ref{oneloop}(a) with $\phi_1$ second order in $k_{\phi W}$). 
Adapting the same gauge-fixing conditions of $k_{\phi\phi}^A$, one can show that the W-propagator with one $k_{\phi W}$ 
inclusion becomes $2im_W^2(k_{\phi W})_{\mu\nu}/g(q^2-m_W^2)^2$. 
Computation of diagrams (Fig. \ref{oneloop}(a)-(i) plus their 
permutations) shows us the $(k_{\phi W})_{\mu\nu}, 
(k_{\phi W})_{\mu\lambda}p^{\lambda}p_{\nu}$, and $(k_{\phi W})_{\lambda\nu}p^{\lambda}p_{\mu}$ structures in the first order 
and $(k_{\phi W})_{\mu\lambda}{(k_{\phi W})^{\lambda}}_{\nu}$ and $(k_{\phi W})_{\mu\lambda}(k_{\phi W})_{
\lambda^{\prime}\nu}p^{\lambda}p^{\lambda^{\prime}}$ in the second order. The only surviving term is the last one which is 
gauge-invariant. Consequently, like the $k_{\phi B}$ case, a very similar relation between $k_F$ and $k_{\phi W}$, 
$(k_F)_{\mu\lambda\lambda^{\prime}\nu}=-\frac{5}{12e^2}\sin^2{\theta_W}(k_{\phi W})_{\mu\lambda}(k_{\phi W})_{\lambda^{\prime}\nu}$, 
yields an upper bound of  $1.7\times 10^{-16}$. It is seen that the current bound on all 
three Lorentz-violating coefficients 
is of the order of $10^{-16}$. 

\section{Coordinate and field redefinitions and the symmetric coefficients}

In this section, we consider bounds on the $k_{\phi\phi}^S$ coefficients.  In this case, the strongest
bounds come from relating, through field redefinitions, these coefficients to other Lorentz violating
coefficients in the fermion sector, and then using previously determined bounds on those coefficients.

Once one extends a model by  relaxing one or more symmetry properties of the original model, the extended model should 
involve  all possible otherwise invariant structures. However, if the modification is carried out under the assumption that
the fields are transformed under this otherwise broken symmetry group in the usual way, not all of new parameters representing 
apparent violation of this symmetry may be physical (i.e. the model has some redundant parameters). 
 Therefore an extension 
should be carefully analyzed to check for redundant parameters. This analysis 
may yield several Lagrangians which are equivalent
to each other by some coordinate and field redefinitions and rescalings \cite{Colladay:1998fq,Colladay:2002eh,Colladay:2003tj,Muller:2004mb}. 
The same situation applies to the SME case.  A simple example is 
provided by Colladay and Kosteleck\'y \cite{Colladay:1998fq}.   Consider the electron in QED, 
with the kinetic term 
$\overline{\psi}\gamma^{\mu}D_{\mu}\psi$.  Suppose one 
transforms the electron field as $\psi\rightarrow exp(-ia^{\mu}x_{\mu})\psi$, where $a$ is a constant vector.  This is not a gauge 
transformation, since $A_{\mu}$ is not changed.  Plugging into the 
kinetic term, one finds a term $a_{\mu}\overline{\psi}\gamma^{\mu}\psi$. 
But this is one of the Lorentz-violating terms mentioned in the first 
section, and thus this term can have no physical effect.   Other field 
redefinitions can eliminate (or, more precisely, make redundant) other 
possible terms.    Recently, the spinor part of the extended
QED has been extensively discussed by Colladay and 
McDonald \cite{Colladay:2002eh}. The $a_{\mu}$ term need not be redundant if gravity is included. This has been explored \cite{Kostelecky:2003fs} by studying the SME with gravity in the context of Riemann-Cartan spacetimes, and thus new Lorentz-violating coefficients appear in such a framework.

In the Higgs sector, one can also make some of the symmetric 
coefficients redundant.  Here we just 
consider the  $U(1)$ part but the generalization to $SU(2)\times U(1)$ 
is straightforward.  A toy model discussed in \cite{kostelecky:2002hh,Muller:2004mb} is relevant 
to our purpose. Consider first a model involving only two Lorentz-violating parameters $k_{\phi\phi}$ and $k_F$ in the scalar and photon 
sectors, respectively. The Lagrangian is ${\cal L}=\left[g_{\mu\nu}+(k_{\phi\phi})_{\mu\nu}\right](D^{\mu}\Phi)^{
\dagger}D^{\nu}\Phi-m^2\Phi^{\dagger}\Phi-\frac{1}{4}F_{\mu\nu}F^{\mu\nu}-\frac{1}{4}(k_F)_{\mu\lambda\lambda^{\prime}\nu}F^{\mu\lambda}F^{\lambda^{
\prime}\nu}$, where $D_{\mu}=\partial_{\mu}+iqA_{\mu}$ and $k_{\phi\phi}$ 
is real and symmetric.  First let us assume that only 
one component of $k_{\phi\phi}$, $(k_{\phi\phi})_{00}\!\equiv\! k^2\!-\!1$, is 
nonzero \cite{kostelecky:2002hh,Muller:2004mb} and that $k_{F}$ is taken as zero.
  By making 
the coordinate transformations $t\to k t$, ${\bf x}\to {\bf x}$ and the field redefinitions $A_0\to A_0$, ${\bf A}\to k{\bf A}$ with 
rescaling of the electric charge $q\to q/k$, one gets the Lagrangian  
${\cal L}_{\rm photon}=(D_{\mu}\Phi)^{\dagger}D^{\mu}\Phi-m^2\Phi^{\dagger}\Phi+\frac{1}{2}(E^2-k^2 B^2)$, where $E(B)$ is the electric(magnetic) field. So, we start with a system having a Lorentz violation in the scalar sector ($k_F=0$) and end up with an equivalent Lagrangian 
involving Lorentz violation in photon sector (some components of $k_F$ are nonzero). 
Second we can further show that by choosing\footnote{This choice was made in Ref. \cite{Iltan:2003nq}, where it was shown that the contribution to Higgs decays from this term is negligible.} 
only 
$(k_{\phi\phi})_{11}=(k_{\phi\phi})_{22}=(k_{\phi\phi})_{33}=\!k^2\!-\!1$ 
nonzero it is still possible to get an equivalent Lagrangian 
as ${\cal L}_{\rm 
photon}=(D_{\mu}\Phi)^{\dagger}D^{\mu}\Phi-m^2\Phi^{\dagger}\Phi+\frac{1}{2}(E^2-B^2/k^2)$ 
under the       
 transformations $t\to t$, ${\bf x}\to k {\bf x}$ and the redefinitions 
$A_0\to k A_0$, ${\bf A}\to {\bf A}$ with the same 
 charge rescaling $q\to q/k$. However, for the other components of  
$k_{\phi\phi}$, there are no such obvious transformations. 

Another analysis of the physical effects of the Lorentz-violating coefficients $k_{\phi\phi}^S$ 
can be found by looking at the effects of field redefinitions over those parameters. 
These effects in the fermion sector were discussed in detail in the context of extended QED \cite{Colladay:2002eh}. 
There it was shown that under the fermion field redefinition $\psi(x)=(1+c_{\mu \nu}x^{\mu}\partial^{\nu})\chi(x)$ it is possible 
to generate a would-be Lorentz-violating Lagrangian in the free fermion context and $c_{\mu\nu}$ represents the Lorentz violation. 
Here $c_{\mu\nu}$ is a real symmetric  coefficient of the Lorentz 
violating $c_{\mu\nu}\bar{\psi}\gamma^{\mu}D^{\nu}\psi$ term in the 
fermion sector. However, 
this transformed Lagrangian can further be expressed in terms of a new
coordinate system having a non-diagonal metric, i.e. a skewed coordinate system, and in this way it is possible to restore the form of 
the original Lagrangian. In this framework, this shows that $c_{\mu \nu}$ is not physical. The redundancy of $c_{\mu\nu}$, however, 
disappears when the fermion-photon interaction is involved. 
A very similar analysis for the scalar sector of a toy model, involving a conventional fermion sector with a scalar field $\phi$, 
gives us ${\cal L}(\psi,\Phi)={\cal L}_{0}^f(\psi)+{\cal 
L}^H_{0}(\varphi)+\left[{1\over 2}(k_{\phi\phi}^S)_{\mu\nu}(\partial^{\mu}\varphi^{\dagger})\partial^{\nu}\varphi +{\rm H.c.}\right]$, 
where the scalar field redefinition 
$\Phi(x)=\left(1+\frac{1}{2}(k_{\phi\phi}^S)_{\mu\nu}x^{\mu}\partial^{\nu}\right)\varphi(x)$ 
is assumed. 
Again expressing the fields in terms of skewed coordinates with a modified metric $\eta_{\mu\nu}=g_{\mu\nu}+(k_{\phi\phi}^S)_{\mu\nu}$
the apparent Lorentz-violating ($k_{\phi\phi}^S$)-term can be absorbed in the scalar sector but it reappears in the fermion sector as a $c$-term. 
If we further extend our model by including fermion-photon interactions one can show that there is a mixing among 
$k_{\phi\phi}^S, c_{\mu\nu}$, and nine unbounded $k_F$ coefficients \cite{privatekost}. 
Consequently, the observability of $k_{\phi\phi}^S$ is nothing but a matter 
of convention. The above analysis enables us to move a non-zero 
$k_{\phi\phi}^{S}$ term into either a $c_{\mu\nu}$ term or a $k_{F}$ 
term. In this Letter we {\it only} concentrate on the Lorentz and CPT violation in the 
scalar sector of the SME, 
hence we assume that the theory has a conventional fermion sector, which 
means that bounds on $c_{\mu\nu}$ will lead to effective bounds on 
$k_{\phi\phi}^{S}$.  A full and systematic analysis of all of the field redefinitions and 
redundancies in the SME would be valuable, but is beyond the scope of this paper.
With our normalizations, a bound on $c_{\mu\nu}$ 
will translate directly into an equivalent bound on 
$(k_{\phi\phi}^{S})_{\mu\nu}$.

We thus need the current bounds on the $c_{\mu\nu}$ coefficients. Although 
numerous bounds appear in the literature, many of them should be taken 
{\it cum grano salis}. Consider the spatial parts of $c_{\mu\nu}$. The 
strongest bounds give an upper limit on the 
diagonal spatial 
elements of $10^{-27}$ \cite{Lamor,Bluhm:2003un,Kostelecky:1999mr} and on the off-diagonal elements $c_{XZ}$ and $c_{YZ}$ of $10^{-25}$ \cite{Prest,Bluhm:2003un,Kostelecky:1999mr}, and $c_{XY}$ of $10^{-27}$ \cite{Lamor,Bluhm:2003un,Kostelecky:1999mr}. There are several caveats, however. First, these are bounds for 
$c_{\mu\nu}$ of the neutron. It is conceivable that the mechanism that 
results in Lorentz violation is proportional to the charge, and these 
experiments would miss the effect. It is also conceivable that a 
version of Schiff's theorem (which shows that in the nonrelativistic 
limit, the electric dipole moment of an atom will vanish, even if it does 
not vanish for constituents) will cause a screening of the $c_{\mu\nu}$ 
coefficients of the quarks. The first effect can be eliminated by 
considering protons or electrons, the second can be eliminated by 
considering electrons. Another caveat is that the bounds on the diagonal 
elements are actually bounds on $c_{XX}-c_{YY}$ and 
$c_{XX}+c_{YY}-2c_{ZZ}$, and thus if the Lorentz violation is isotropic, 
the bounds will not apply. In this case, the vanishing trace condition 
will (as in the case of the double-traceless condition on $k_{F}$) yield, 
when the fermion field is properly normalized, a nonzero $c_{TT}$, and 
thus the bounds on the diagonal spatial elements will be that of the bound 
on $c_{TT}$.

The bound on $c_{TT}$ can be obtained by comparing antiproton cyclotron
frequencies with those of a hydrogen ion
\cite{Gabrielse:1998ee} and a very weak bound of $4\times 10^{-13}$ is extracted.  An 
interesting connection between the dispersion relation for fermions and the $c_{TT}$ coefficient
has been noted by Bertolami, et al.\cite{bertolami}, and astrophysical experiments to improve the bound
is proposed.
For the time-space components,  there are
various studies based on
the sensitivities of some planned experiments
\cite{Bluhm:2003un,kostelecky:2003cr,kostelecky:2003xn,Datta:2003dg};
most
of the bounds are from the neutrino sector of the SME and the highest
proposed sensitivity is around $10^{-25}$
\cite{kostelecky:2003cr}.

\section{Bounds on the CPT-odd coefficient}
The remaining  part of the  Higgs sector Lagrangian has one term that violates both Lorentz and CPT symmetries, represented 
by the complex constant coefficient $(k_{\phi})^{\mu}$. One interesting effect of this term is the modification of the conventional 
electroweak $\rm SU(2)\times U(1)$ symmetry breaking. Minimization of the static potential yields a nonzero expectation value for 
$Z_{\mu}$ boson field of the form $\langle 
Z_{\mu}\rangle_{0}={\sin{2\theta_W}\over q} {\rm Re} (k_{\phi})_{\mu}$. Here we have assumed all the other Lorentz-violating 
coefficients zero.  The nonzero expectation value for the $Z$ will, 
when plugged into the conventional fermion-fermion-$Z$ interaction, yield a 
$b_{\mu}\overline{\psi}\gamma^{\mu}\gamma_{5}\psi$ term.   
Alternatively, one can look at the one-loop effects on the photon 
propagator, however this will yield much weaker bounds. 
 By assuming $k_{\phi}$ is the only Lorentz-violating term in the Higgs 
 sector, one finds that the effective
 $b_{\mu}=\frac{1}{4}Re(k_{\phi})_{\mu}$. If we look at the best current bounds on $b_{\mu}$, from testing of cosmic spatial isotropy for 
 polarized electrons \cite{Hou:2003}, $b_{X,Y}^e\leq 3.1\times10^{-29}$ GeV and $b_Z^e\leq 7.1\times10^{-28}$ GeV in the Sun-centered frame. 
 The best bound comes from the neutron with the use of a two-species noble-gas maser \cite{Bear:2000cd} and it is of 
 the order of $b_{X,Y}^n\leq 10^{-32}$ GeV. Note that in order to get this bound there are some assumption about the nuclear 
 configurations, which make the bound uncertain accuracy to within one or two orders of magnitude.  The bound on the time component 
 of $b_{\mu}$ is around $b_T^n\leq 10^{-27}$ GeV \cite{Cane:2003wp}. 
Therefore, the  best bounds for the real part of 
$(k_{\phi})_{\mu}$ are $10^{-31}$ GeV and $10^{-27}$ GeV for the $X,Y$ 
and 
for the $Z,T$ components, respectively.  The imaginary part of $k_{\phi}$ 
is 
unphysical, since this term in the Lagrangian is a total divergence.

\section{Conclusion}
In this work we have studied the bounds on the Lorentz and/or CPT violating coefficients in the Higgs sector of 
the SME. It is shown that all antisymmetric CPT-even Lorentz-violating coefficients  give 
second-order contributions to the photon vacuum polarization at one-loop. By  comparing with the 
$k_{F}$-term and assuming one of them nonzero in each case (without 
high-precision cancellation), we find  
 $(k_{\phi\phi}^A)_{\mu\nu}, (k_{\phi B})_{\mu\nu}, (k_{\phi W})_{\mu\nu} \lesssim 10^{-16}$. For the symmetric part 
of $k_{\phi\phi}$, after discussing the close connections with the Lorentz-violating coefficients $c_{\mu\nu}$ in the 
fermion sector by means of coordinate and field redefinitions, we conclude that the bounds could 
be determined directly from the $c_{\mu\nu}$-term. In a very similar way we obtain the bound on the CPT
and Lorentz-violating coefficient $(k_{\phi})_{\mu}$ by comparing with $b_{\mu}$-term in the fermion sector. 
The existence of $k_{\phi}$-term leads to a nonzero vacuum value for $Z_{\mu}$ which further enables us to relate $(k_{\phi})_{\mu}$ 
with  $b_{\mu}$ and we find an upper bound of $10^{-31} (10^{-27})$ GeV 
for $X,Y (T,Z)$ components of $(k_{\phi})_{\mu}$. 
Table I lists all the bounds together with their sources.

\begin{table}[t]
\vskip -0.1cm
	\caption{Estimated upper bounds for the Lorentz and CPT violating coefficients in the Higgs sector of the SME.} \label{Table}     
\begin{center}
\begin{minipage}{16cm}

    \begin{tabular}{ccccc}
    \hline\hline
Parameters  &&$\;$ Sources & &$\;\;$Comments\\ \hline
	 &$\;\;\;\;\;\;\;\;$$\tilde{\kappa}_{e^{+}},\tilde{\kappa}_{o^{-}}$    & $c_{\mu\nu}$ & $b_{\mu}$ (GeV)&  \\ \hline 
    $(k_{\phi\phi}^A)_{\mu\nu}$    & $\;\;\;\;\;\;\;\;3\times 10^{-16}$      &    -   &   -  &             -   \\
    $(k_{\phi B})_{\mu\nu}$    & $\;\;\;\;\;\;0.9\times  10^{-16}$    &    -  &   - &               -   \\
    $(k_{\phi W})_{\mu\nu}$   & $\;\;\;\;\;\;\;\;1.7\times 10^{-16}$     &    - &  -  &                   -   \\
    $(k_{\phi\phi}^S)_{II}$   & $\;\;\;\;\;\;$ -    &   $10^{-27}$&  - &  \footnote{Obtained from $c_{\mu\nu}^{\rm neutron}$ with the assumption that Lorentz violation is not isotropic. If it is isotropic,\linebreak {\vskip -1.0cm the bound on $(k_{\phi\phi}^S)_{TT}$ applies.}}                                                                       \\
    $ (k_{\phi\phi}^S)_{TT}$  & $\;\;\;\;\;\;$ - &$4\times 10^{-13}$& - &   \footnote{Obtained from the comparison of the anti-proton's frequency with the hydrogen ion's frequency.}                                                                         \\
    $(k_{\phi\phi}^S)_{TI}$    & $\;\;\;\;\;\;$ -  &    $10^{-25}$& - &  \footnote{Estimated value based on the sensitivity calculations of some planned 
space-experiments.}                                                                        \\
    $(k_{\phi\phi}^S)_{XZ},(k_{\phi\phi}^S)_{YZ}$   & $\;\;\;\;\;\;$ -  &    $10^{-25}$&  - & \footnote{Obtained from the neutron.}                              \\
$(k_{\phi\phi}^S)_{XY}$  & $\;\;\;\;\;\;$ - & $10^{-27}$ & - &  $^{d}$  \\
    $(k_{\phi})_X, (k_{\phi})_Y$   & $\;\;\;\;\;\;$ -   &    -  &   $10^{-31}$ & \footnote{From $b_{\mu}^{\rm neutron}$ with the use of a two-species noble-gas maser. From  $b_{\mu}^{\rm electron}$, a weaker but cleaner bound\linebreak {\vskip -1.0cm of $1.2\times 10^{-25}$ can be obtained.}}          \\
    $(k_{\phi})_Z,(k_{\phi})_T$   & $\;\;\;\;\;\;$ -  &    -  &   $2.8\times 10^{-27}$  &   \footnote{This bound is from the spatial isotropy test of polarized electrons.}                                 \\
    \hline \hline
        \end{tabular}
\end{minipage}

        \end{center}
\end{table}

Perhaps the most intriguing bounds are for the antisymmetric 
coefficients.  Recent developments in string theory indicate that 
Lorentz-violating non-commutative geometry might be a low-energy 
probe of Planck scale physics \cite{Carroll:2001ws,Colladay:2003tj}, and this geometry will be 
antisymmetric. 
It is interesting 
that our upper bounds 
on the coefficients are $O(10^{-16})$, which is less than an order of 
magnitude above the ratio of the electroweak to Planck scale.  An 
improvement in the birefringence bounds of a couple of orders of 
magnitude (which is feasible \cite{kostelecky:2001mb,Costa}) could probe this 
sensitivity.  Should a $k_{F}$ term actually be discovered, our 
analysis shows how one can distinguish Higgs sector Lorentz violation 
from other sectors. Specifically, of the ten observable $k_{F}$ 
coefficients, we find nonzero values only for the two independent 
diagonal elements of $\tilde{\kappa}_{e+}$.  Thus, the origin of 
Lorentz violation might be experimentally accessible. It should be noted that inclusion of gravity might lead to new Lorentz-violating terms, as discussed in Ref. \cite{Kostelecky:2003fs}.

If the primary effects of an underlying Lorentz and CPT violation appear in the Higgs sector, what are the most promising experiments? We have seen that CPT violation will be manifested through a vacuum expectation value of the $Z$ boson, and the ``$b$'' coefficient for a fermion will be proportional to the weak axial coupling of that fermion. Testing this would require $b_f$ to be measured for at least two fermions. For antisymmetric CPT-even Lorentz violation, there are very specific signatures, discussed in the previous paragraph, and improvement in the birefringence bounds of a couple of orders of magnitude would be valuable. For symmetric CPT-even Lorentz violation, there are tight bounds, but with various assumptions and caveats. The relatively weak $c_{TT}$ and $c_{TI}$ bounds, as noted in Ref. \cite{Bluhm:2003un}, could be substantially tightened.
\begin{acknowledgments}

We are extremely grateful to Alan Kosteleck\'y for many useful 
discussions, much advice and encouragement.  We also thank Chris 
Carone for useful discussions.
The work of DA and MS was supported by
        the National Science Foundation under grant PHY-0243400, and the work of IT was supported by 
      the Scientific and Technical Research Council of Turkey (T\"{U}B\.ITAK) in the framework of
      NATO-B1 program.
\end{acknowledgments}


\end{document}